\documentclass[10pt,a4paper,showpacs,superscriptaddress,twocolumn]{revtex4-1}
\usepackage{amssymb}
\usepackage{graphicx}
\usepackage{amsmath}
\usepackage{subfigure}
\usepackage{float}

\newcommand{\bear}{\begin{eqnarray}}
\newcommand{\eear}{\end{eqnarray}}
\newcommand{\be}{\begin{equation}}
\newcommand{\ee}{\end{equation}}
\newcommand{\beqn}{\begin{eqnarray}}
\newcommand{\eeqn}{\end{eqnarray}}
\newcommand{\beqnn}{\begin{eqnarray*}}
\newcommand{\eeqnn}{\end{eqnarray*}}

\begin{document}

\title{Controlling the nonclassical properties of a hybrid Cooper pair box
system and an intensity dependent nanomechanical resonator.}
\author{C.~Valverde}
\email{valverde@ueg.br}
\affiliation{C\^{a}mpus Henrique Santillo, Universidade
 Estadual de Goi\'{a}s, Rod. BR 153, km 98, 75.132-903 An\'{a}polis, GO, Brazil}
\affiliation{Universidade Paulista (UNIP), Rod. BR 153, km 7, 74.845-090
Goi\^ania, GO, Brazil} 
\affiliation{Escola Superior de Neg\'{o}cios-ESUP, Av. Ant\^{o}nio Fid\'{e}lis 515,
74.840-090 Goi\^{a}nia, GO, Brazil}

\author{V.G.~Gon\c{c}alves}
\affiliation{Universidade Paulista (UNIP), Rod. BR 153, km 7, 74.845-090
Goi\^ania, GO, Brazil}

\author{B.~Baseia}
\affiliation{Instituto de F\'{\i}sica, Universidade Federal de Goi\'as,
74.001-970 Goi\^ania, GO, Brazil}
\date{\today }

\begin{abstract}
We employ a more realistic treatment to investigate the entropy and the
excitation-inversion of a coupled system that consists of a nanomechanical
resonator and a superconducting Cooper pair box. The procedure uses the
Buck-Sukumar model in the microwave domain, considers the nanoresonator with
a time dependent frequency and both subsystems in the presence of losses.
Interesting results were found for the temporal evolutions of the entropy of
each subsystem and of the excitation-inversion in the Cooper pair box. A
comparison was also performed about which of these two subsystems is more
sensitive to the presence of losses. The results suggest that appropriate
choices of the involved time dependent parameters allow us to monitor these
two features of the subsystems and may offer potential applications, e.g.,
in the generation of nonclassical states, quantum communication, quantum
lithography.
\end{abstract}

\pacs{42.65.Yj; 03.65.Yz; 42.50.Nn}
\maketitle

\section{Introduction}

In the last years the investigations on nanomechanical systems \cite{ca1}
have rapidly been developed. The rush in this direction was estimulated by
various perspectives and applications before unsuspected. The enormous
progress in the research of nanomechanical systems has also placed the dream
of controlling the interface between the quantum and classic worlds in a
realistic way. This was shown by the emergence of hybrid quantum systems 
\cite{ca2,ca3}, which are intended to achieve a coherent transfer of quantum
information from a single quantum emitter (e.g., superconducting qubits,
cooper pair box, microwave resonators, quantum dots, etc) and a solid state
mechanical resonator, which moves in quantum physics giving birth to a new
paradigm. In addition, these systems have received considerable attention
due to their diverse potential applications, including metrology (mass force
or spin ultra-sensitive detectors), based on the remarkable sensing
properties of nanomechanical resonators and their very low mass and
lightness.\ 

An important focus of quantum optics is concerned with the atom-photon
system. Inspired on the various tests applied to this coupled system and on
the several results obtained, including the limitations, the researchers
have passed from the light domain to the microwave domain of the
superconducting version, the quantum electrodynamics circuit. This system
furnishes a new test for microwave \textquotedblleft \textit{photons}%
\textquotedblright\ that interact with superconducting qubits\cite{y1,n1}.\
In this scenario the atom is substituted by a Cooper pair box (\textbf{CPB})
while the photon is substituted by a nanomechanical resonator (\textbf{NR}%
).\ So, the \textit{atom-field} interacting system goes to the \textbf{%
\textbf{CPB }- \textbf{NR}} interacting system with the concomitant passage
from the optical domain to the microwave domain.

There are few works in the literature that treat the interaction between a 
\textbf{CPB} and a \textbf{NR} when the later has either a time dependent
frequency \cite{15,15a,15a2,15a4} or a time dependent amplitude of
oscillation \cite{16,16a}. The well known Jaynes-Cummings model ($JCM$),
which describes the interaction of a single two-level atom and a single mode
of a quantized radiation field, is the simplest model for this system and
provides exact solutions. Analogous to the \textbf{\textbf{CPB }- \textbf{NR}%
} system, but since the year 1963, many others studies have previously
implemented the $JCM$ to describe the atom-field interaction \cite%
{15a2,15a4,15a1,15a3,15a5}. Some generalized models were also constructed
and extensively studied \cite{b2,b3,b4,b5}. The examples include the study
of the mentioned systems in presence of the Stark effect\cite{r1,w1}, e.g.,
to investigate quantum nondemolition measurements \cite{b6,b7,b8,b9}.
Usually the investigations assume the field initially in a (pure) coherent
state. As the atom-field interacion is turned on, the field state changes
with time the evolution. Then, if its coherence is lost, the field state
becomes nonclassical \cite{1a,1a1,1a2}. Some references to this subject are,
e.g., on nonclassical properties of a state \cite{1b,1b1,1b2}; generation of
superposition states\cite{1b3,1b4}; on the degree of nonclassicality of a
state \cite{1c} and: sculpturing coherent states to get Fock states \cite{1d}
- plus references therein.

The traditional $JCM$ was extended to the case of \textit{%
intensity-dependent coupling}, proposed by B. Buck and C.V. Sukumar \cite%
{buck} in order to study the influence of the field intensity, via its
excitation number, upon the atom-field system, where a single atom interacts
with a single mode of an optical field. This model was generalized by V.
Buzek \cite{buzek} to include a new coupling, with time dependent intensity.
\ In the present work we will employ the model by Buck-Sukumar ($BS$) to
study the \textbf{\textbf{CPB }- \textbf{NR}} system, namely: we will
suppose the coupling being dependent of the intensity of the NR oscillations
and also that it changes with time, as assumed in \cite{buzek}. In addition,
we consider a more realistic scenario with the presence of dissipation
effects and verify in which way they affect the excited level of the \textbf{%
CPB}$,$ the excitations of the \textbf{NR}$,$ and the dynamical properties
of the entire \textbf{\textbf{CPB }- \textbf{NR}} system. Some points
considered here are: how dissipation spoils the system operation and in
which way the detuning could prevent it, allowing us the control of
entanglement features, collapse-revival effects, and others. The results
obtained indicate the possibility of some potential applications \cite%
{aa1,aa2,aa3,aa4,aa5,aa6}.

\section{The Hamiltonian System}

\begin{figure*}[tbh]
\centering  
\fbox{\includegraphics[width=8cm, height=6cm]{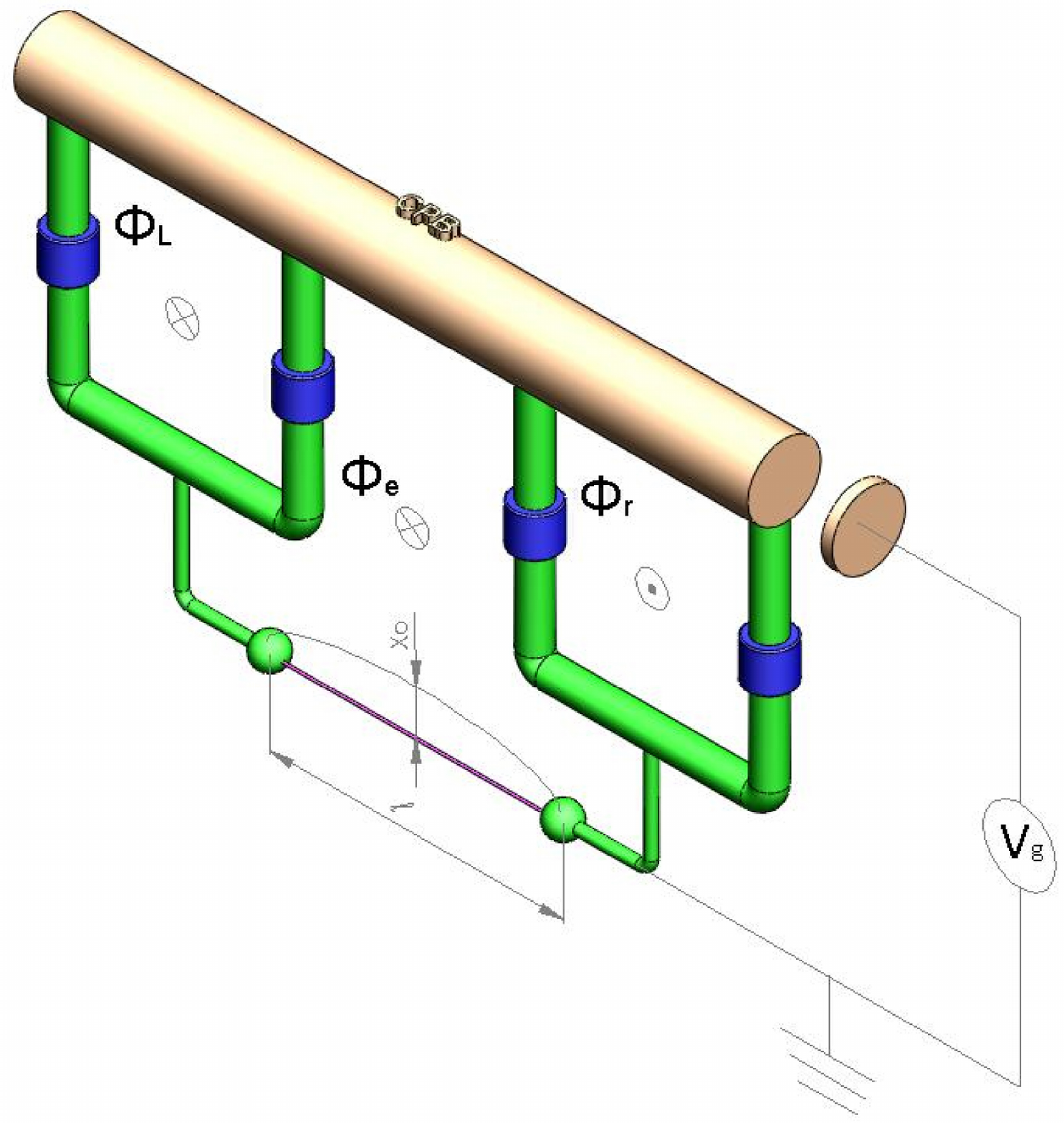}}
\caption{Model of a Cooper pair box system coupled to a nanoresonator.}
\label{cooper}
\end{figure*}

A superconductor \textbf{CPB} charge qubit is adjusted\ to the input voltage 
$V_{1}$ of the system, through a capacitor with an input capacitance $C_{1}$%
. Following the configuration shown in Fig. \ref{cooper} we observe three
loops: a small loop in the left, another in the right, and a great loop in
the center. The control of the external parameters of the system can be
implemented via the input voltage $V_{1}$ and the three external fluxes $%
\Phi _{L},$ $\Phi _{r}$ and $\Phi _{t}$.\ The control of theses parameters
allows us to make the coupling between the \textbf{CPB} and the \textbf{NR}.
We consider $\hslash =1$ and assume as identical the four Josephson
junctions of the circuit system, having the same Josephson energy $E_{J}^{0}$%
; the external fluxes $\Phi _{L}$\ and $\Phi _{r}$\ are also assumed as
identical in magnitude, although they have opposite signs $\Phi _{L}=-\Phi
_{r}=\Phi _{x}$ (see Ref. \cite{15a2}).\ So, taking into account the decay
in the excited level of the \textbf{CPB} and dissipation in the \textbf{NR}$%
, $\ we can write the total Hamiltonian of the system as follows,

\begin{widetext}
\begin{equation}
\hat{H}=\omega (t)\hat{a}^{\dagger }\hat{a}+\frac{1}{2}\omega _{c}(t)\hat{%
\sigma}_{z}+\lambda (t)\left( \hat{a}\sqrt{\hat{a}^{\dagger }\hat{a}}\hat{%
\sigma}_{+}+\sqrt{\hat{a}^{\dagger }\hat{a}}\hat{a}^{\dagger }\hat{\sigma}%
_{-}\right) -i\gamma (t)\left\vert e\rangle \langle e\right\vert -i\delta (t)%
\hat{a}^{\dagger }\hat{a}.  \label{a1}
\end{equation}%
\end{widetext}

In the Eq. (\ref{a1}) the first term describes the \textbf{NR}, the second
describes the \textbf{CPB}, the third represents the intensity dependent
interaction introduced by the $BS$ model, here including the time dependent
parameter $\lambda (t)$ introduced by Buzek, the fourth term $\gamma (t)$
stands for the time dependent loss affecting the \textbf{CPB} and the fifth
term $\delta (t)$ stands for the time dependent loss that affects the 
\textbf{NR}. In the Eq. (\ref{a1}) $\hat{a}^{\dagger }$ ($\hat{a}$) is the
creation (anihilation) operator for excitations in the \textbf{NR}, $\hat{%
\sigma}_{+}$ ($\hat{\sigma}_{-}$) is the raising (lowering) operator for the 
\textbf{CPB}, and $\hat{\sigma}_{z}$ is the z-component of the Pauli spin
operator. In the Hamiltonian of Eq. (\ref{a1}) stands for a non Hermitian
Hamiltonian ($NHH$). Usually, quantum mechanics works with hermitian
Hamiltonians; however, one can found a lot of papers in the literature using 
$NHH:$ one of them appears in Ref. \cite{nhh1}, used by the authors to map a
wave guide system, metal-silicon, where an optical potential is modulated
along the length of the wave guide to get a non reciprocal light
propagation; another application appears in Ref. \cite{nhh}, employed by the
authors to show occurrence of entanglement in many body systems; in the Ref. 
\cite{nh10} the author study optical realizations of a relativistic $NHH$;
in Ref. \cite{nh5} the authors use a $NHH$ and a convenient algoritm to
generalize a conventional theory; in Ref. \cite{nh1} a $NHH$ is applied to
get information about input-output channels; an approach in Ref. \cite{nh6}
neglects the hermiticity to study canonical transformations in quantum
mechanics; Ref. \cite{nh7}\ applies a $NHH$ to solve a quantum master
equation; Ref. \cite{nh3} uses a $NHH$ in a new approach to study a weak
spectral density of H-bonds and attenuations in parts of the system; Ref. 
\cite{nh8}, studies a $NHH$ to show that it exhibits real eigenvalues; Ref. 
\cite{nh4} employs a canonical formulation to study a dissipative quantum
mechanics that exhibits complex eigenvalues; Ref. \cite{nh9} investigates a $%
NHH$ in non commutative spaces; in the Ref. \cite{l22} a $NHH$ is considered
to include damping in a system described by $JCM$ and to study entangled
states of atoms in distinct cavities; etc. \ 

In addition to the previous references concerned with $NHH,$ here it is
pertinent remembering an alternative treatment by Carl M. Bender \cite%
{bender}, who defended the efficacy of the non hermitian $PT$-symmetric
Hamiltonians. Accordingly, although the hermiticity of the Hamiltonian is
sufficient to guarantee the two essential properties of quantum mechanics
(namely, reality of eigenvalues and unitary of the time evolution), it is
not necessary. Actually, various complex Hamiltonians are also able to
guarantee these two properties and they include the real symmetric
Hamiltonian as a special case, required by the $PT$-symmetric quantum
mechanics. (Indeed,\ even the supposed guarantee assigned to hermitian
operators is controversial \cite{ja0,ja00}.) Although \ the theoretical
approach in this line began around 1998, the first experimental results came
in the last five years \cite{ja,jb,jc,jd,je,jf}.

In the Hamiltonian of Eq.(\ref{a1}) $\hat{\sigma}_{+}=\left\vert e\rangle
\langle g\right\vert ,$ $\hat{\sigma}_{-}=\left\vert g\rangle \langle
e\right\vert $ and $\hat{\sigma}_{z}=\left\vert e\rangle \langle
e\right\vert -\left\vert g\rangle \langle g\right\vert $\ stand respectively
for the transition $(\hat{\sigma}_{+},\hat{\sigma}_{-})$ and atomic
inversion $(\hat{\sigma}_{z})$ operators; they act in the space of atomic
states with frequency $\omega _{c}(t)$; $\ \hat{a}^{\dagger }$ and $\hat{a}$
are the (non hermitian) creation and annihillation operators of the \textbf{%
NR} excitations, with frequency $\omega \left( t\right) =\omega _{0}+f(t)$: $%
f(t)$ is a $t$-dependent function, which can also be constant; $\gamma (t)$\
is the decay parameter of the CPB, from its excited level $\left\vert
e\right\rangle $\ to its fundamental level $\left\vert g\right\rangle $; $%
\delta (t)$ is the decay parameter of the \textbf{NR} and$\ \lambda (t)$ is
the coupling coefficient between the \textbf{CPB} and the \textbf{NR}.

As known in the literature the coupling parameter $\lambda (t)$\ can be
written in the form\ \cite{18},

\begin{equation}
\lambda (t)=|\vec{d}_{ij}|\sqrt{\frac{\omega \left( t\right) }{2\epsilon
_{0}V\left( t\right) }},  \label{b3}
\end{equation}%
where the quantization volume $V\left( t\right) $ is dependent on time and
takes the form $V\left( t\right) =\frac{V_{0}}{1+f\left( t\right) /\omega
_{0}}$; $\epsilon _{0}$ is the permissivity constant; \ $\vec{d}%
_{ij}=e\left\langle i\right\vert \vec{r}\left\vert j\right\rangle $ \ is the
matrix element of the dipole between the two \textbf{CPB}\ states $%
\left\vert i\right\rangle \ $and$\ \left\vert j\right\rangle ;$ $e$ is the
elementary charge, $\vec{r}$ is the position vector and, $\left\vert
i\right\rangle \ $and$\ \left\vert j\right\rangle $ play similar roles of
the atomic states $\left\vert e\right\rangle \ $and$\ \left\vert
g\right\rangle $ in the scenario of atom-field interaction. The notation $%
|e\rangle ,$ $|g\rangle $ will be used from now on. It stands respectively
for the excited and ground states of the \textbf{CPB}.

Substituting the expressions of the field frequency $\omega \left( t\right) $
and quantum volume $V\left( t\right) $ in Eq. (\ref{b3}), we can write it as

\begin{equation}
\lambda (t)=\lambda _{0}\sqrt{1+\frac{f(t)}{\omega _{0}}},  \label{bb4}
\end{equation}%
where $\lambda _{0}=|\vec{d}_{eg}|\sqrt{\frac{\omega _{0}}{2\epsilon
_{0}V_{0}}}$\emph{.} The control of the parameters $\omega (t)$\ and $%
\lambda (t)$\ is provided by an external field that acts upon the \textbf{NR}%
\ and changes the magnetic flux $\Phi _{e}$\ (cf. Fig. \ref{cooper}).

The wave function that describes the \textbf{\textbf{CPB }- \textbf{NR}}
system as function of the time $t$ can be written in the form,

\begin{equation}
\left\vert \Psi \left( t\right) \right\rangle =\sum_{n}^{\infty }\left[
C_{e,n}\left( t\right) \left\vert e,n\right\rangle +C_{g,n}\left( t\right)
\left\vert g,n\right\rangle \right] ,  \label{a2}
\end{equation}%
where the coefficients $C_{e,n}\left( t\right) $\ and $C_{g,n}\left(
t\right) $\ stand respectively for the probability amplitudes to find the
entire system in the states $\left\vert e,n\right\rangle $ and $\left\vert
g,n\right\rangle $. This notation represents the \textbf{CPB} in its excited
state $\left\vert e\right\rangle $ and fundamental state $\left\vert
g\right\rangle $ with $n$ excitations\ in the \textbf{NR}. We will assume
the subsystems \textbf{CPB} and the \textbf{NR} are decoupled at $t=0$ with
the \textbf{CPB} in its excited state $\left\vert e\right\rangle $ and the 
\textbf{NR} prepared in a coherent state $\left\vert \alpha \right\rangle ,$
this later expressed by the superposition of Fock states,

\begin{equation}
\left\vert \alpha \right\rangle =\sum\limits_{n=0}^{\infty }F_{n}\left\vert
n\right\rangle ,  \label{cb1}
\end{equation}%
where $F_{n}=\frac{\alpha ^{n}}{\sqrt{n!}}e^{-\left\vert \alpha \right\vert
^{2}/2}.$ In this way the wave function of the whole system in the the
initial state can be written as $\left\vert \Psi (0)\right\rangle
=\left\vert e\right\rangle \left\vert \alpha \right\rangle =\sum_{n}^{\infty
}F_{n}\left\vert e,n\right\rangle ,$ since\emph{\ }the initial conditions
restrict the probability amplitudes to $C_{g,n}(0)=0$ for all values of\emph{%
\ }$n=0,1,2,3,...$ and $\sum_{n=0}^{\infty }\left\vert C_{e,n}(0)\right\vert
^{2}=1.$

By analyzing the time evolution of the \textbf{\textbf{CPB }- \textbf{NR}}
system, described by the time\ dependent Schr\"{o}dinger equation,

\begin{equation}
i\hbar \frac{d\left\vert \Psi \left( t\right) \right\rangle }{dt}=\hat{H}%
\left\vert \Psi \left( t\right) \right\rangle ,  \label{a3}
\end{equation}%
and the Hamiltonian $\hat{H}$\ given in the Eq.(\ref{a1}), we find the set
of equations of motion for the\ coefficients $C_{e,n}\left( t\right) $ and $%
C_{g,n+1}\left( t\right) $,

\begin{widetext}
\begin{equation}
\frac{\partial C_{e,n}\left( t\right) }{\partial t}=\left( -in\omega(t) -i\frac{%
\omega _{c}(t)}{2}-\gamma (t)-n\delta (t)\right) C_{e,n}\left( t\right)
-i\lambda (t)(n+1)C_{g,n+1}\left( t\right) ,  \label{a9}
\end{equation}

\begin{equation}
\frac{\partial C_{g,n+1}\left( t\right) }{\partial t}=\left( -i(n+1)\omega(t) +i%
\frac{\omega _{c}(t)}{2}-(n+1)\delta (t)\right) C_{g,n+1}\left( t\right)
-i\lambda (t)(n+1)C_{e,n}\left( t\right) .  \label{a10}
\end{equation}%
\end{widetext}

Solving the set of equations (\ref{a9}) and (\ref{a10}) we obtain the
solutions for $C_{e,n}(t)$ and $C_{g,n+1}(t)$. To this end, we have used the
Runge-Kutta method of $4^{th}$\ order; \ with theses solutions we determine
the quantum dynamical properties of the system, including the entanglement
that affects the subsystems \textbf{CPB} and \textbf{NR}.

\section{Entropy of the \textbf{CPB }- \textbf{NR} System}

Recently, several authors have employed various methods to investigate the
dynamics of entanglement \cite{15a1,l3v,k41,k6}. Here the name
\textquotedblleft entanglement\textquotedblright\ means \textquotedblleft
mixing of states\textquotedblright , whose degree can be measured by the
entropy. The Von Neumann entropy offers a quantitative measure of the system
disorder or the degree of a quantum state purity, e.g., as shown by Phoenix
and Knight \cite{19}. Here the entropy defined in the form, 
\begin{equation}
S_{\mathbf{NR}(\mathbf{CPB})}=-Tr_{N(C)}(\hat{\rho}_{N(C)}\ln \hat{\rho}%
_{N(C)}),
\end{equation}%
is a measure related to the entanglement of one of the two interacting
subsystems, with $\hat{\rho}_{N(C)}=Tr_{C(N)}(\hat{\rho}_{NC})$ where $\hat{%
\rho}_{N(C)}$\ is the density operator that describes the state of the
subsystem \textbf{NR}$($\textbf{CPB}$)$ and $\hat{\rho}_{NC}$ is the same
operator for the entire system \textbf{\textbf{CPB }- \textbf{NR}}. The
quantum dynamics represented by the Hamiltonian in Eq. (\ref{a1}) creates
the entanglement in both subsystems of the \textbf{\textbf{CPB }- \textbf{NR}%
}. In the following discussion we use the quantum entropy of Von Neumann as
a measure of the degree of entanglement. For a quantum system with two
components the entropy obeys the Araki-Lieb theorem, which states that $%
\left\vert S_{\mathbf{CPB}}-S_{\mathbf{NR}}\right\vert \leq S\leq S_{\mathbf{%
CPB}}+S_{\mathbf{NR}}$. One consequence of this inequality is that, if the
total system is initially prepared in a pure state, then the entropies of
the components of the system remain with equal value in their subsequent
time evolution. Then it is sufficient to study one of them to know both.

So, assuming the two subsystems in pure states at $t=0$\ the entropies of
the \textbf{CPB}\ and \textbf{NR}\ subsystems, which are found via the $BS$\
model, are identical: $S_{\mathbf{CPB}}(t)=S_{\mathbf{NR}}(t).$\ For
example, the entropy of the \textbf{NR}\ is found from the equation,%
\begin{equation}
S_{\mathbf{NR}}(t)=-\left[ S_{\mathbf{NR}}^{+}(t)\ln \left( S_{\mathbf{NR}%
}^{+}(t)\right) +S_{\mathbf{NR}}^{-}(t)\ln \left( S_{\mathbf{NR}%
}^{-}(t)\right) \right]
\end{equation}%
where :%
\begin{widetext}
\begin{eqnarray}
S_{NR}^{\pm }(t) &=&\frac{1}{2}\Bigg(\sum_{n=0}^{\infty }\left\vert
C_{e,n}(t)\right\vert ^{2}+\sum_{n=0}^{\infty }\left\vert
C_{g,n+1}(t)\right\vert ^{2}\pm \frac{1}{2}\Bigg[\Bigg(\sum_{n=0}^{\infty
}\left\vert C_{e,n}(t)\right\vert ^{2}-\sum_{n=0}^{\infty }\left\vert
C_{g,n+1}(t)\right\vert ^{2}\Bigg)^{2}  \notag \\
&&+4\left\vert \sum_{n=0}^{\infty }C_{e,n+1}^{\ast
}(t)C_{g,n+1}(t)\right\vert ^{2}\Bigg]^{1/2}\Bigg).
\end{eqnarray}%
\end{widetext}

The results obtained from the calculations are shown in the Figs. (\ref{s1}, %
\ref{s12}, \ref{s2} and \ref{s3})

\section{The \textbf{CPB} Excitation Inversion}

The \textbf{CPB} excitation inversion$,$ denoted by $I(t),$ is an important
observable of the two-level systems. It is defined as the difference of
probabilities of finding the \textbf{CPB} in the excited and fundamental
states. The mathematical expression representing this property is, \ 
\begin{equation}
I(t)=\sum\limits_{n=0}^{\infty }\left[ \left\vert C_{e,n}(t)\right\vert
^{2}-\left\vert C_{g,n+1}(t)\right\vert ^{2}\right] .  \label{aa1}
\end{equation}%
The results achieved from the Eq. (\ref{aa1}), for various values of
parameters, are shown in Figs. (\ref{inv1} and \ref{inv2}).

\section{Results and \ Discussions}

\begin{figure}[h!tb]
\centering
\subfigure(a){\includegraphics[height=4cm]{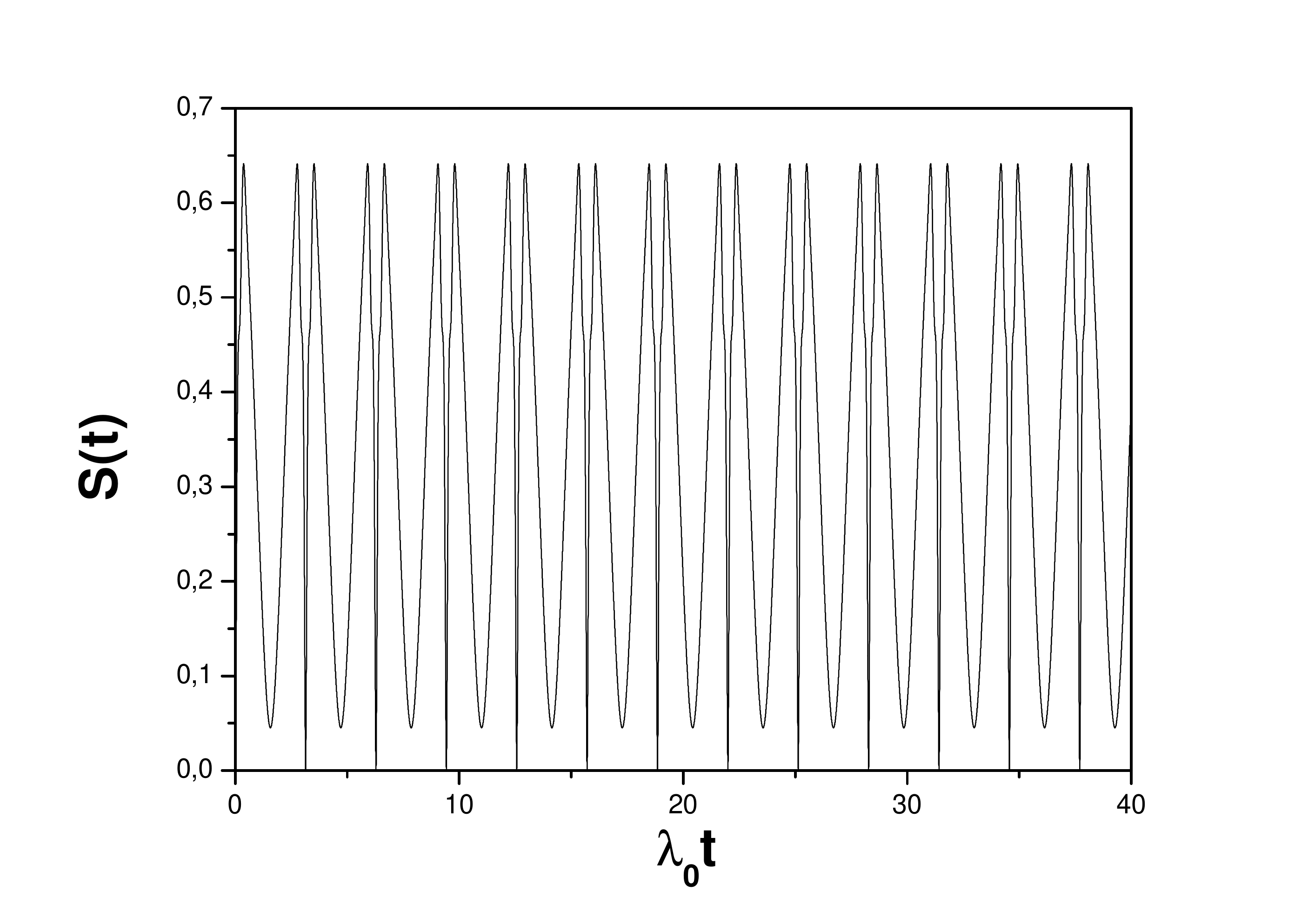} \label{s1a} } \quad 
\subfigure(b){\includegraphics[height=4cm]{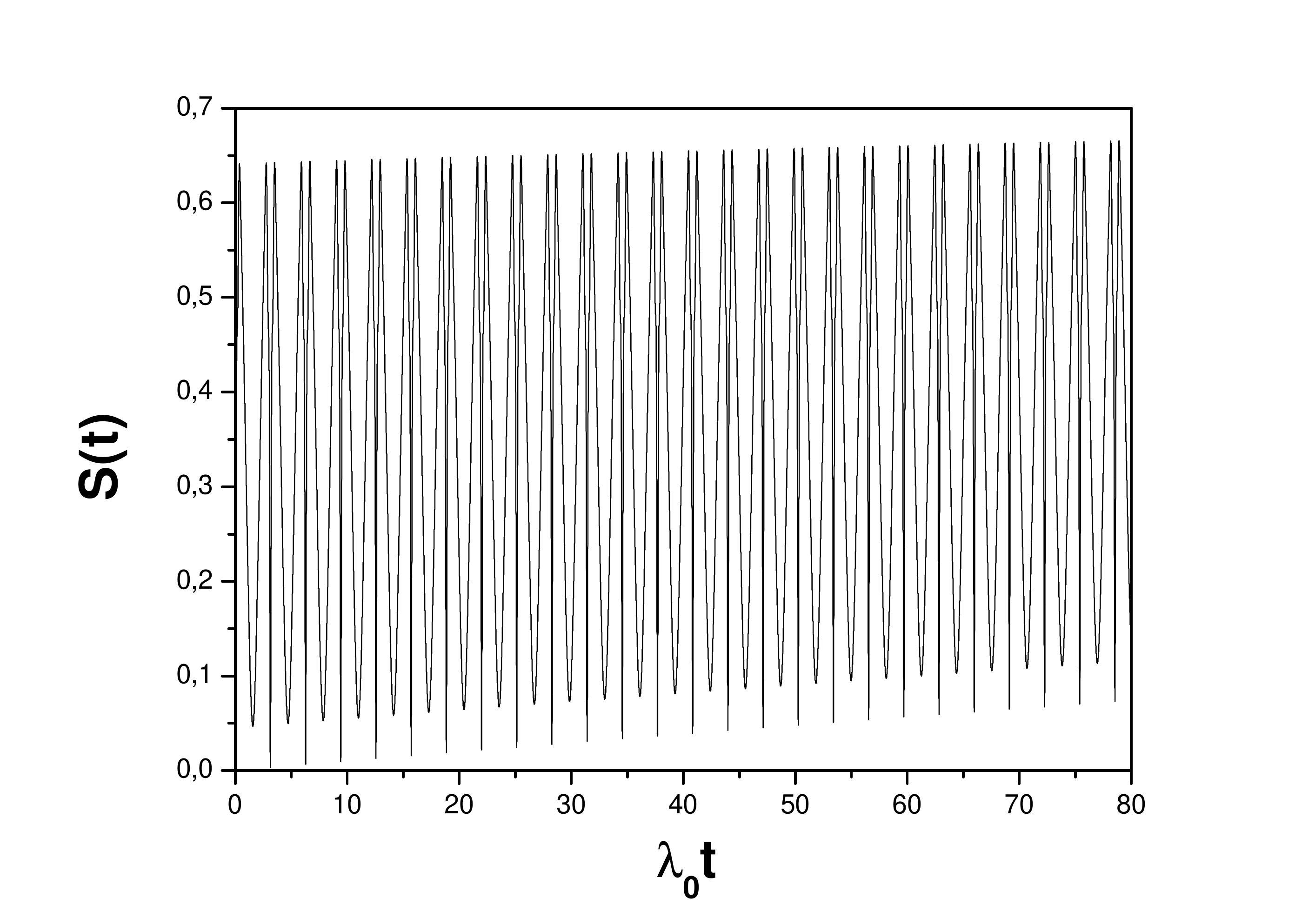} \label{s1b} }
\caption{Time evolution of the entropy for different values of the
parameters $\protect\gamma (t)$ and $\protect\delta (t)$ for: $\left\langle
n\right\rangle =9$, $\protect\omega _{0}=\protect\omega _{c}=2000\protect%
\lambda _{0}$, $f(t)=0$. (a) $\protect\gamma =0,0\protect\lambda _{0}$\ and $%
\protect\delta =0.0\protect\lambda _{0}$; (b) $\protect\gamma =0,001\protect%
\lambda _{0}$\ and $\protect\delta =0.0\protect\lambda _{0}$.}
\label{s1}
\end{figure}

Firstly, for $f(t)=0$ we will consider the time evolution of the \textbf{NR}
entropy for various values of the decay coefficients,\ $\gamma (t)$ and $%
\delta (t).$ We assume the \textbf{NR} initially in a coherent state\ $%
|\alpha \rangle $\ with average number of excitations $\left\langle \hat{n}%
\right\rangle =|\alpha |^{2}=9$ and the resonant case $\omega _{0}=\omega
_{_{C}}=2000\lambda _{0};$ other values of parameters are used in Fig. \ref%
{s1}. In an ideal system the parameters $\gamma (t)$ and $\delta (t)$ are
null, as used in Fig. \ref{s1} (a). In this figure, the maximum of the 
\textbf{NR} entropy is close to $\ln 2$. After the start of the interaction
the \textbf{NR} entropy gradually goes to its minimum, then returns to its
maximum and remains oscillating regularly.

\begin{figure}[h!tb]
\centering
\subfigure(a){\includegraphics[height=4cm]{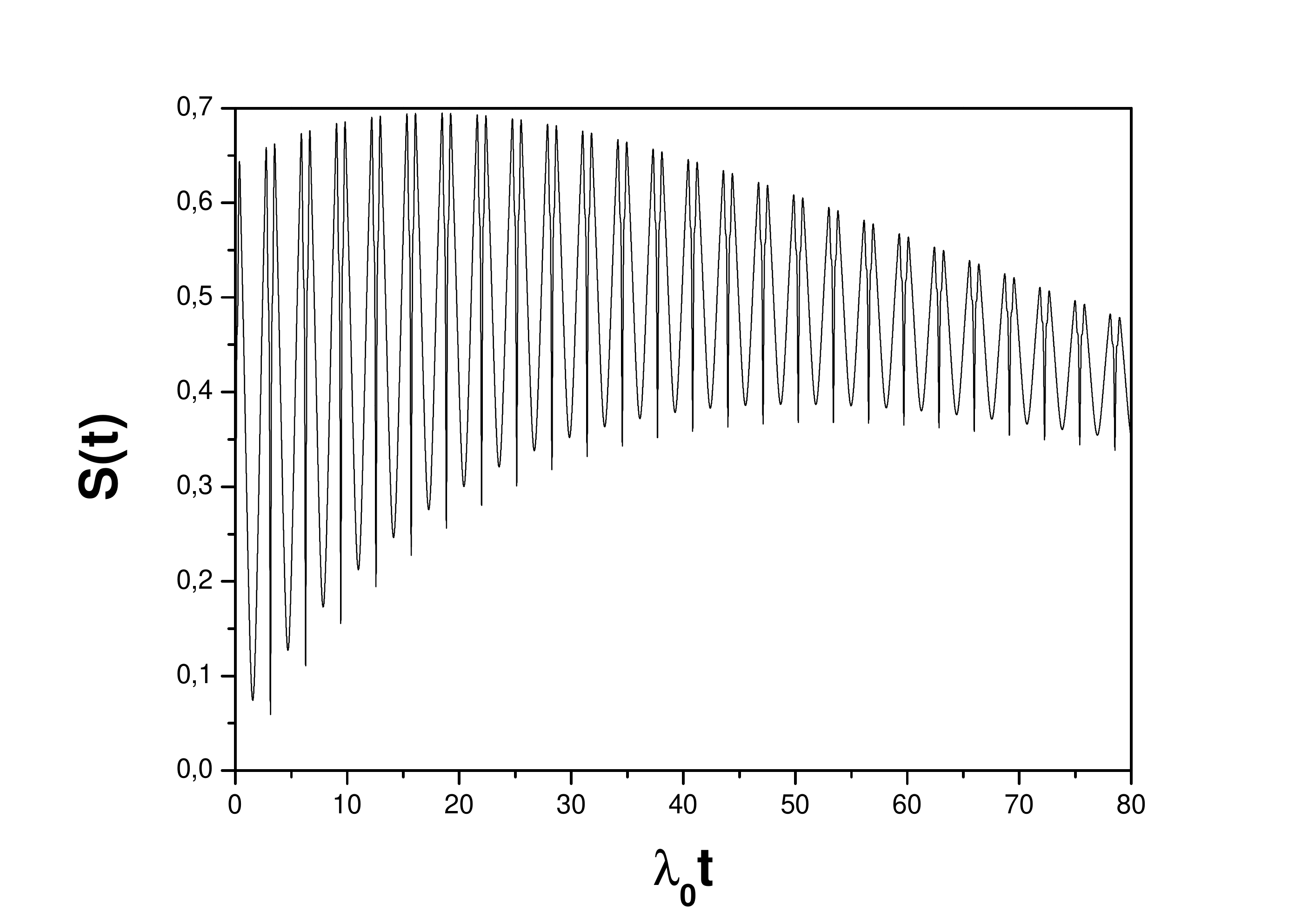} \label{s12c} } \quad 
\subfigure(b){\includegraphics[height=4cm]{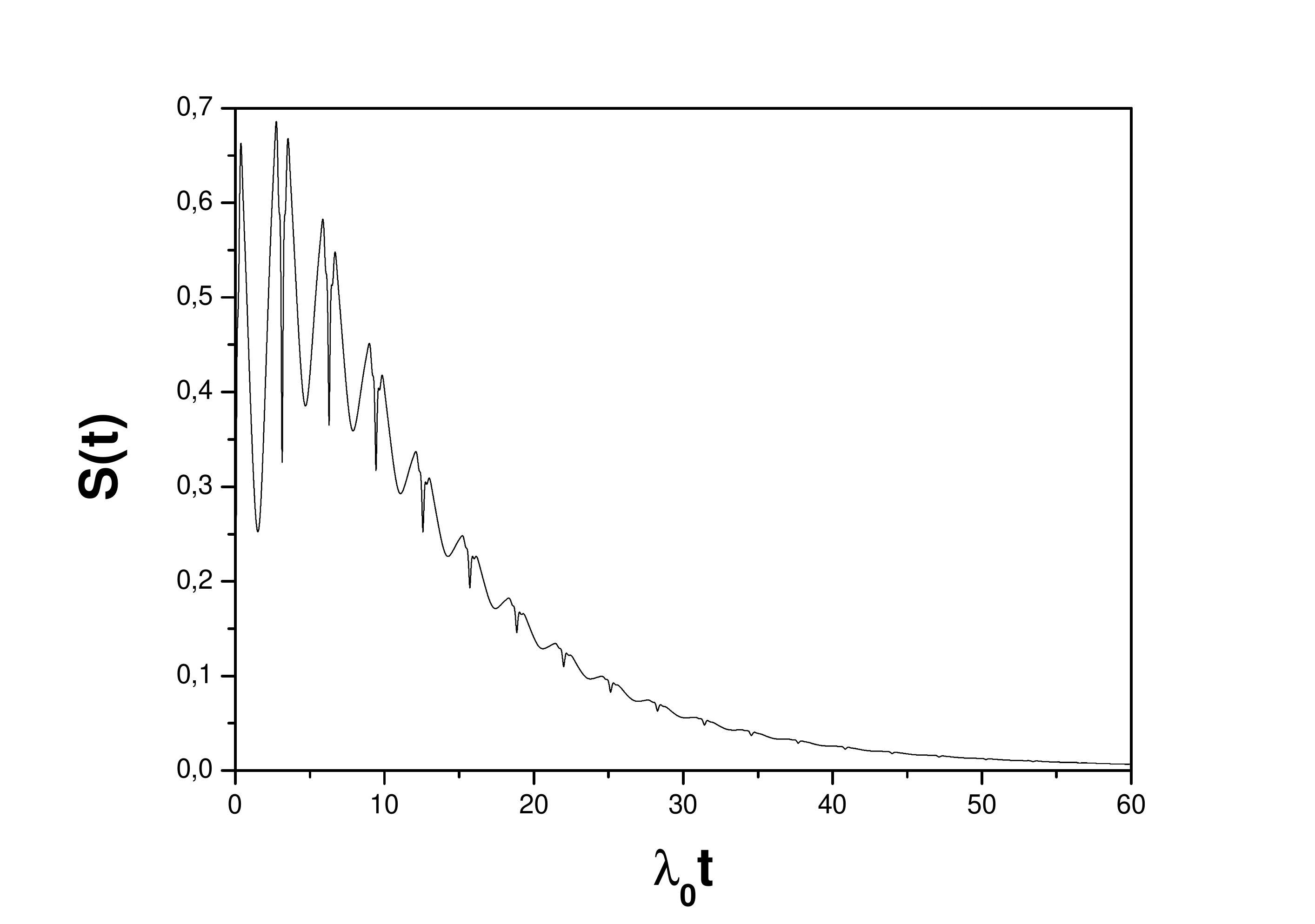} \label{s12d} }
\caption{Time evolution of the entropy for different values of the
parameters $\protect\gamma (t)$ and $\protect\delta (t)$ for: $\left\langle
n\right\rangle =9$, $\protect\omega _{0}=\protect\omega _{c}=2000\protect%
\lambda _{0}$, $f(t)=0$. (a) $\protect\gamma =0.001\protect\lambda _{0}$\
and $\protect\delta =0.001\protect\lambda _{0}$; (b) $\protect\gamma =0.001%
\protect\lambda _{0}$ and $\protect\delta =0.01\protect\lambda _{0}$.}
\label{s12}
\end{figure}

For small values of decay in the \textbf{CPB}$,$ as $\gamma =0.001\lambda
_{0},$ and an ideal \textbf{NR} $(\delta =0)$ the maximum value of the
entropy shows no significative changes for small times (cf. Fig. \ref{s1}
(b)). For larger values of time the amplitude of the entropy oscillations
diminishes while the entropy itself moves away from the value $S(t)=0$. If
instead we include a small decay in the \textbf{NR}, as $\delta
=0.001\lambda _{0},$ the entropy changes as follows: although the amplitude
of the entropy oscillations again decreases, now the entropy moves slowly to
its minimum value $S(t)=0$ (cf. Fig. \ref{s12} (a)). For larger value of $%
\delta (t)$ the entropy movies rapidly to zero, which is due to the passage
of both subsystems to their respective ground states. In this case the
entropy loses its periodicity (cf. Fig. \ref{s12} (b)).

\begin{figure}[h!tb]
\centering
\subfigure(a){\includegraphics[height=4cm]{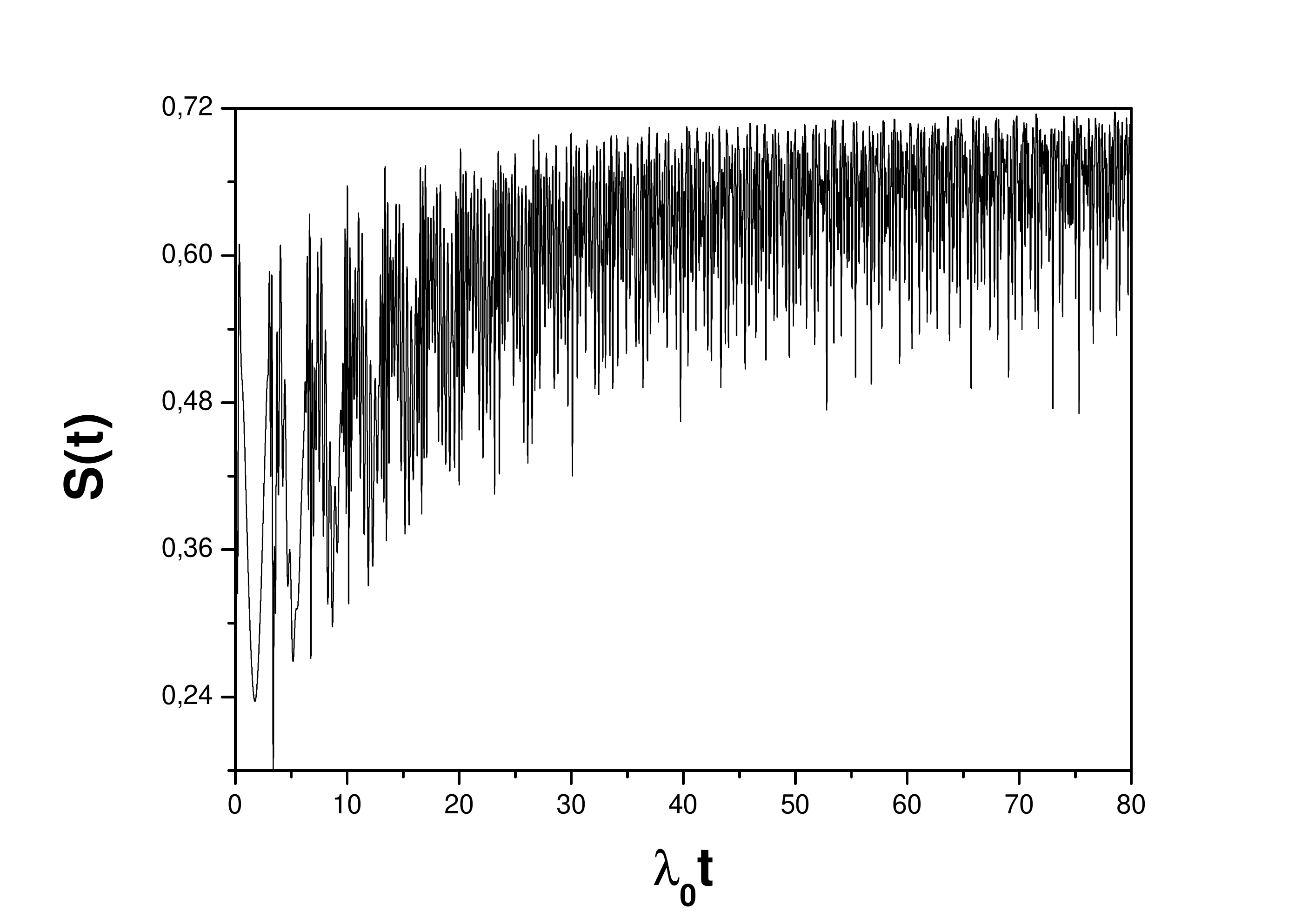} \label{s2a} } \quad 
\subfigure(b){\includegraphics[height=4cm]{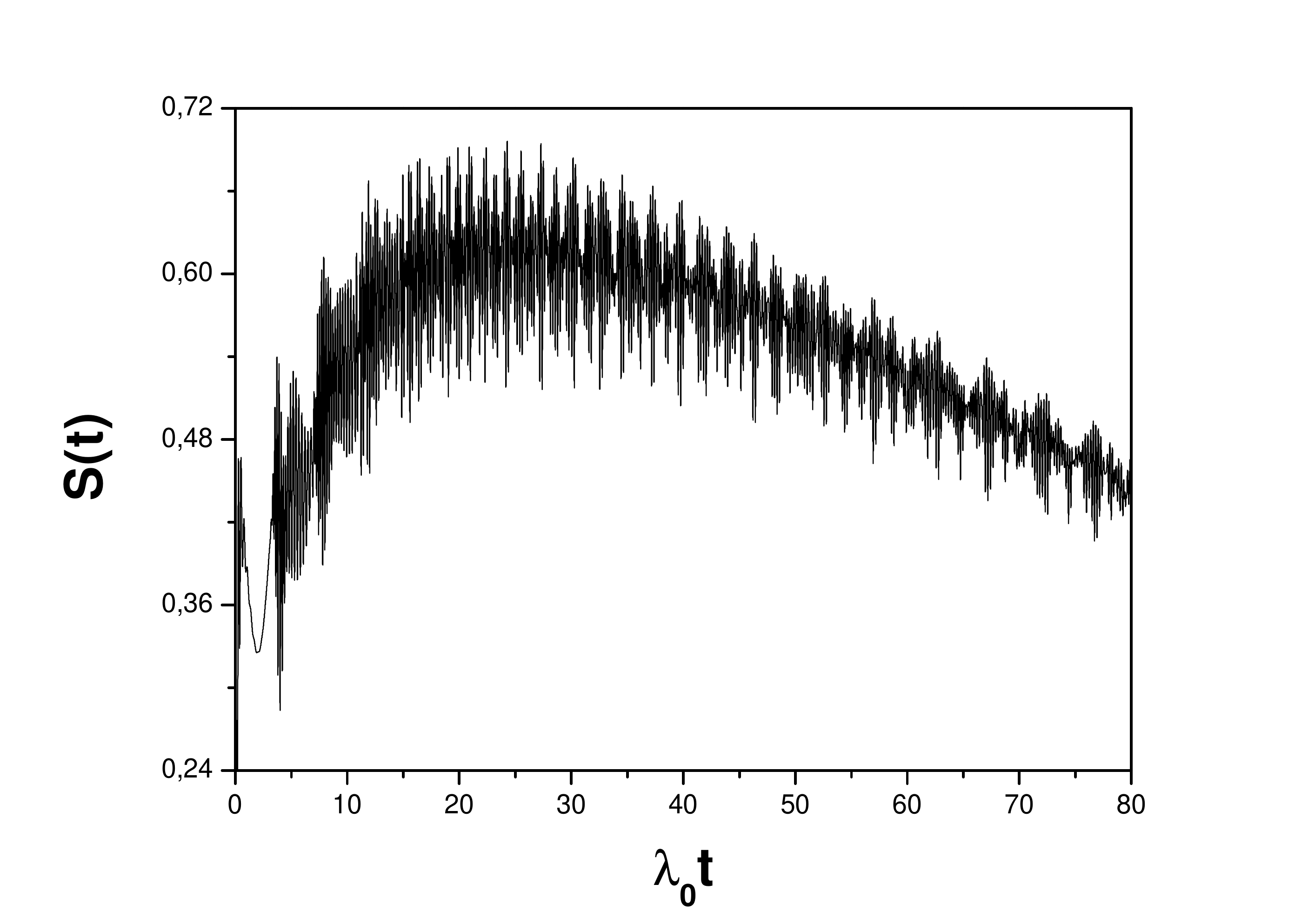} \label{s2b} }
\caption{Time evolution of the entropy for different values of the
parameters $\protect\gamma (t)$ and $\protect\delta (t)$ for: $\left\langle
n\right\rangle =9$, $\protect\omega _{0}=\protect\omega _{c}=2000\protect%
\lambda _{0}$ (a) $\protect\gamma =0.001\protect\lambda _{0}$\ and $\protect%
\delta =0.0\protect\lambda _{0}$, $f(t)=10\protect\lambda _{0};$ (b) $%
\protect\gamma =0.001\protect\lambda _{0}$\ and $\protect\delta =0.001%
\protect\lambda _{0}$, $f(t)=20\protect\lambda _{0}.$}
\label{s2}
\end{figure}

Here we consider the nonresonant case, namely, for $f(t)=\Delta =const$,
with $\Delta \ll \omega _{c}$, $\omega _{0}$. Comparing the Fig. \ref{s2}
(a) with Fig. \ref{s1} (b), both have the same values of the decay
coefficients $\gamma (t)$ and $\delta (t)$,\ we observe the amplitude of
entanglement decreasing as the detuning $\Delta $ increases, i.e., the
detuning turns the entropy oscillation greater while destroying its
periodicity. In this scenario $(\Delta \neq 0)$ if the \textbf{CPB} decay
parameter $\gamma (t)$ increases the entropy goes quickly to zero (not shown
in figures). Comparing the Fig. \ref{s2} (b) with Fig. \ref{s12} (a) we see
a similar behavior of the average values of the entropies whereas the
amplitude of oscillations diminishes in the case $\Delta \neq 0.$ All these
comparisons show that the influence of detuning upon the \textbf{NR} entropy
is greater in the ideal \textbf{NR} $(\delta =0)$. In the resonant case,
when the decay parameters $\gamma (t)$ and $\delta (t)$ increase the \textbf{%
NR} entropy tends to zero. \ As a consequence of the mentioned Araki-Lieb
theorem, the same occurs in the \textbf{CPB}. 

\begin{figure}[h!tb]
\centering
\subfigure(a){\includegraphics[height=4cm]{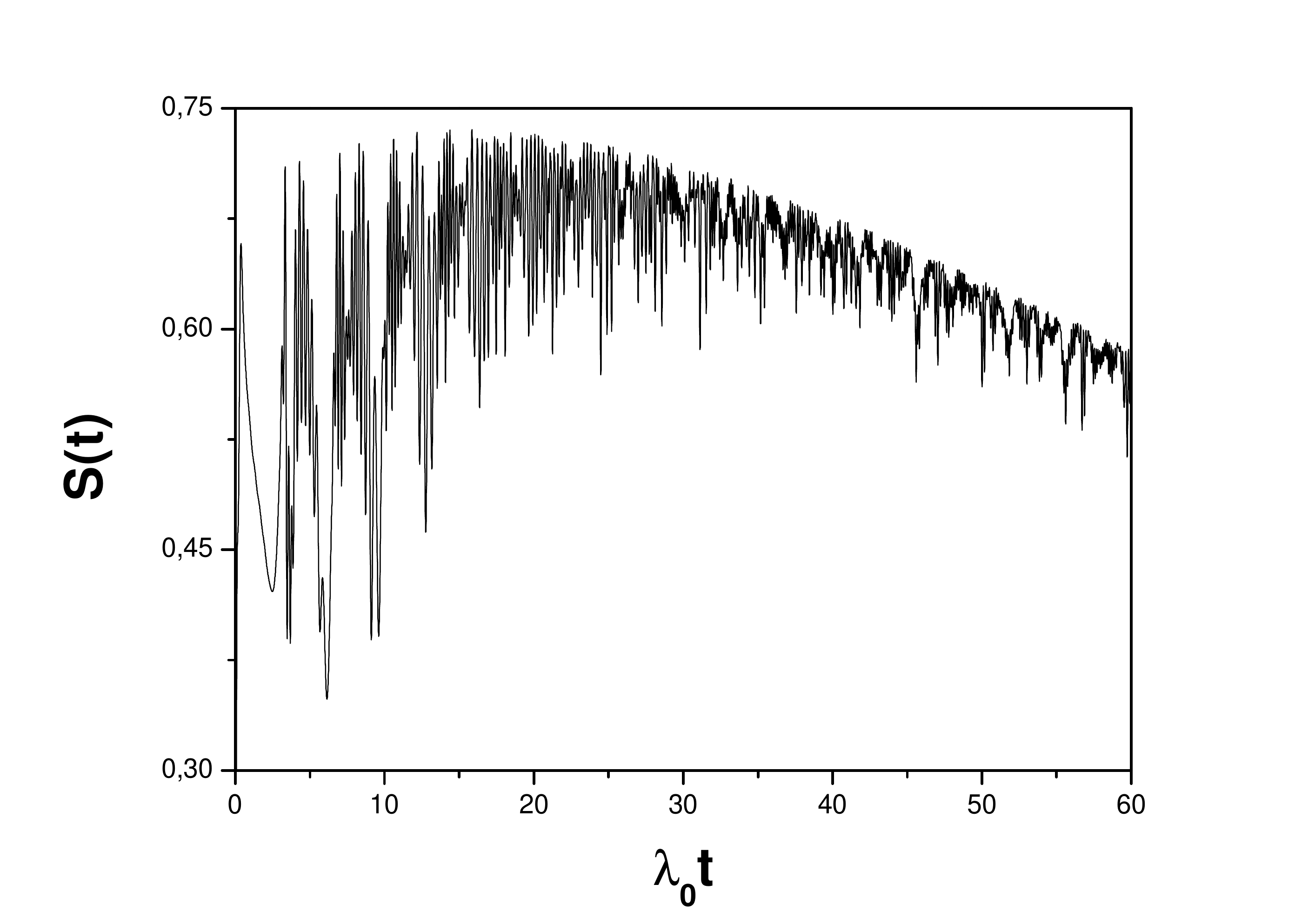} \label{s3a} } \quad 
\subfigure(b){\includegraphics[height=4cm]{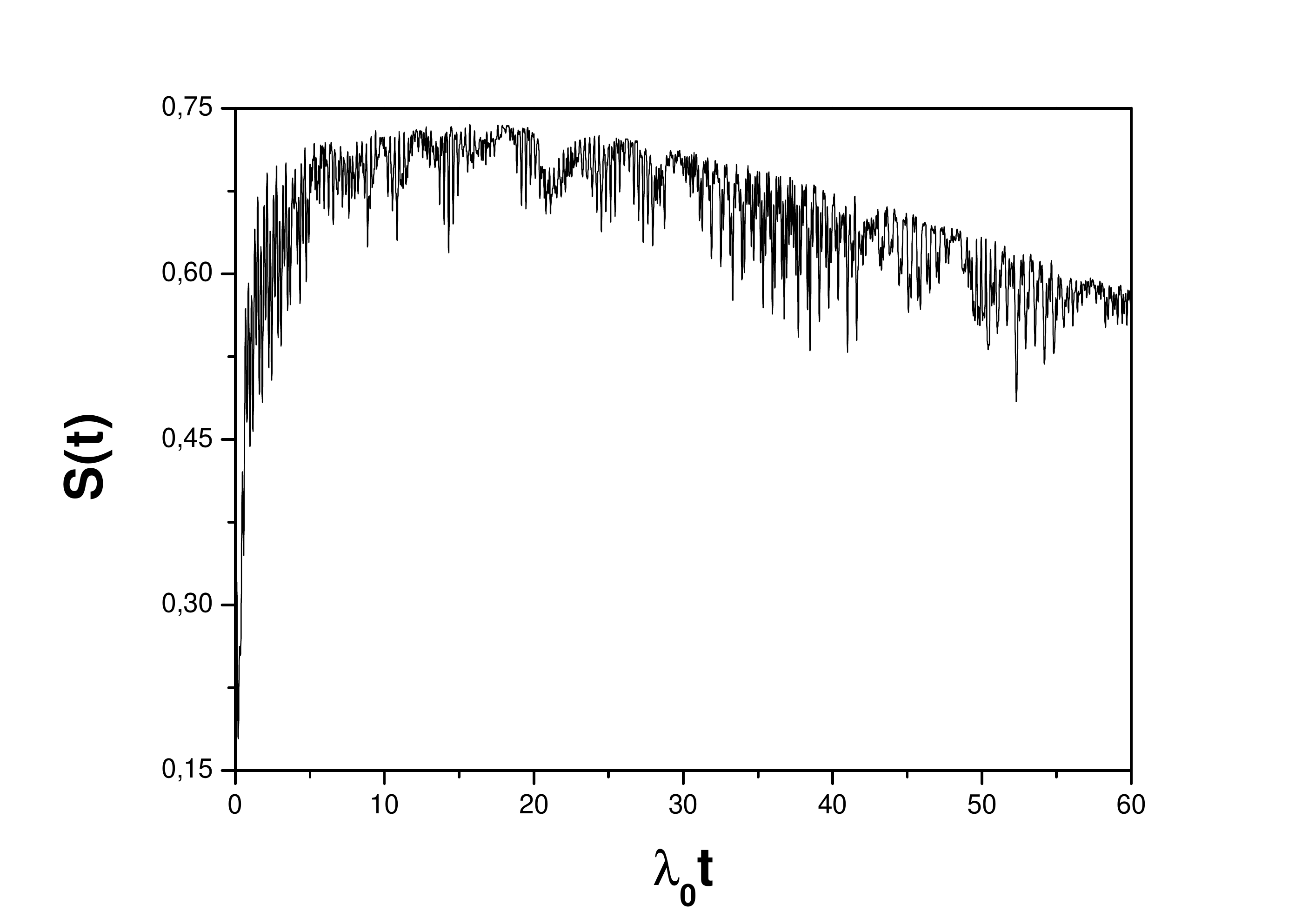} \label{s3b} }
\caption{Time evolution of the entropy for different values of the
parameters $\protect\gamma (t)$ and $\protect\delta (t)$ for: $\left\langle
n\right\rangle =9$, $\protect\omega _{0}=\protect\omega _{c}=2000\protect%
\lambda _{0}$, $\protect\gamma =0.001\protect\lambda _{0}$\ and $\protect%
\delta =0.001\protect\lambda _{0}$; (a) $\protect\eta =20\protect\lambda %
_{0} $, $\protect\omega \prime =\protect\lambda _{0}$; (b) $\protect\eta =20%
\protect\lambda _{0}$, $\protect\omega \prime =20\protect\lambda _{0}$.}
\label{s3}
\end{figure}

Let us now consider the variation in the detuning parameter $\Delta $: \ we
fist take $\Delta \neq 0$ and $f(t)=\eta \sin (\omega \prime t)$, where $%
\eta $ and $\omega \prime $ are parameters that modulates the \textbf{NR}
frequency. Our discussion is limited to the condition $\eta \ll \omega _{c}$%
, $\omega _{0}$ and also assuming that $\omega \prime $\ is small to avoid
interaction of the \textbf{CPB} with other modes of the \textbf{NR}. We have
chosen various values of amplitude modulations $\eta $ to verify the
entanglement properties between the \textbf{CPB} and \textbf{NR}. We use
various values of frequency modulation\ $\omega \prime $ to see its
influence upon the \textbf{\textbf{CPB }- \textbf{NR}} entanglement (cf.
Figs. \ref{s3}).

Comparing the Fig. \ref{s2} (b) with the Figs. \ref{s3} (a), \ref{s3} (b) we
note the entropy exhibiting periodic and quasi periodic oscillations: when
the parameter $\eta $ increases the amplitude of these oscillations
diminishes. This means that the modulation of the \textbf{NR} sinusoidal
frequency\ is important to stabilize entanglements in the \textbf{\textbf{%
CPB }- \textbf{NR}}\ system.

\begin{figure}[h!tb]
\centering
\subfigure(a){\includegraphics[height=4cm]{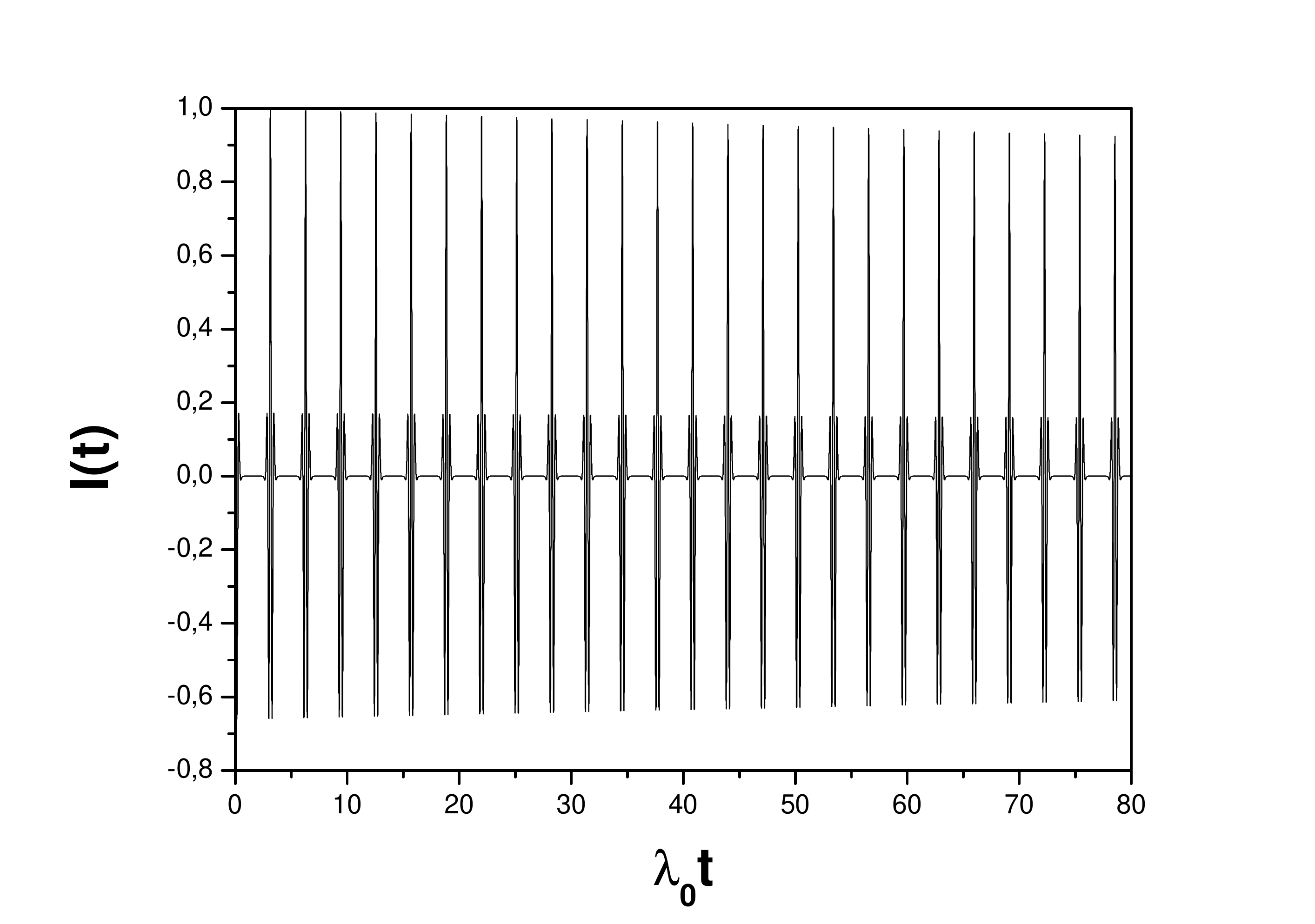} \label{i1a} } \quad 
\subfigure(b){\includegraphics[height=4cm]{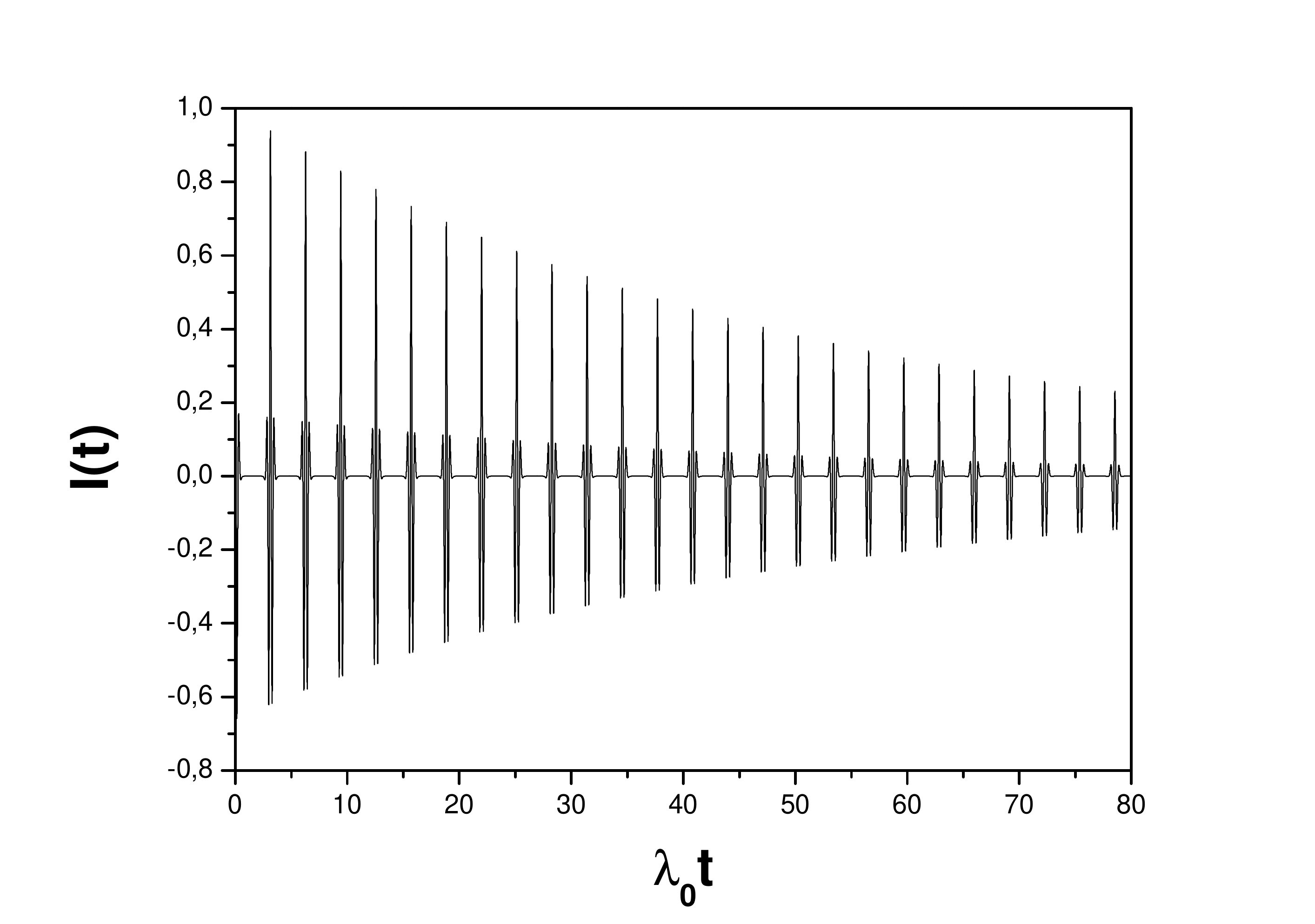} \label{i1b} } \quad 
\subfigure(c){\includegraphics[height=4cm]{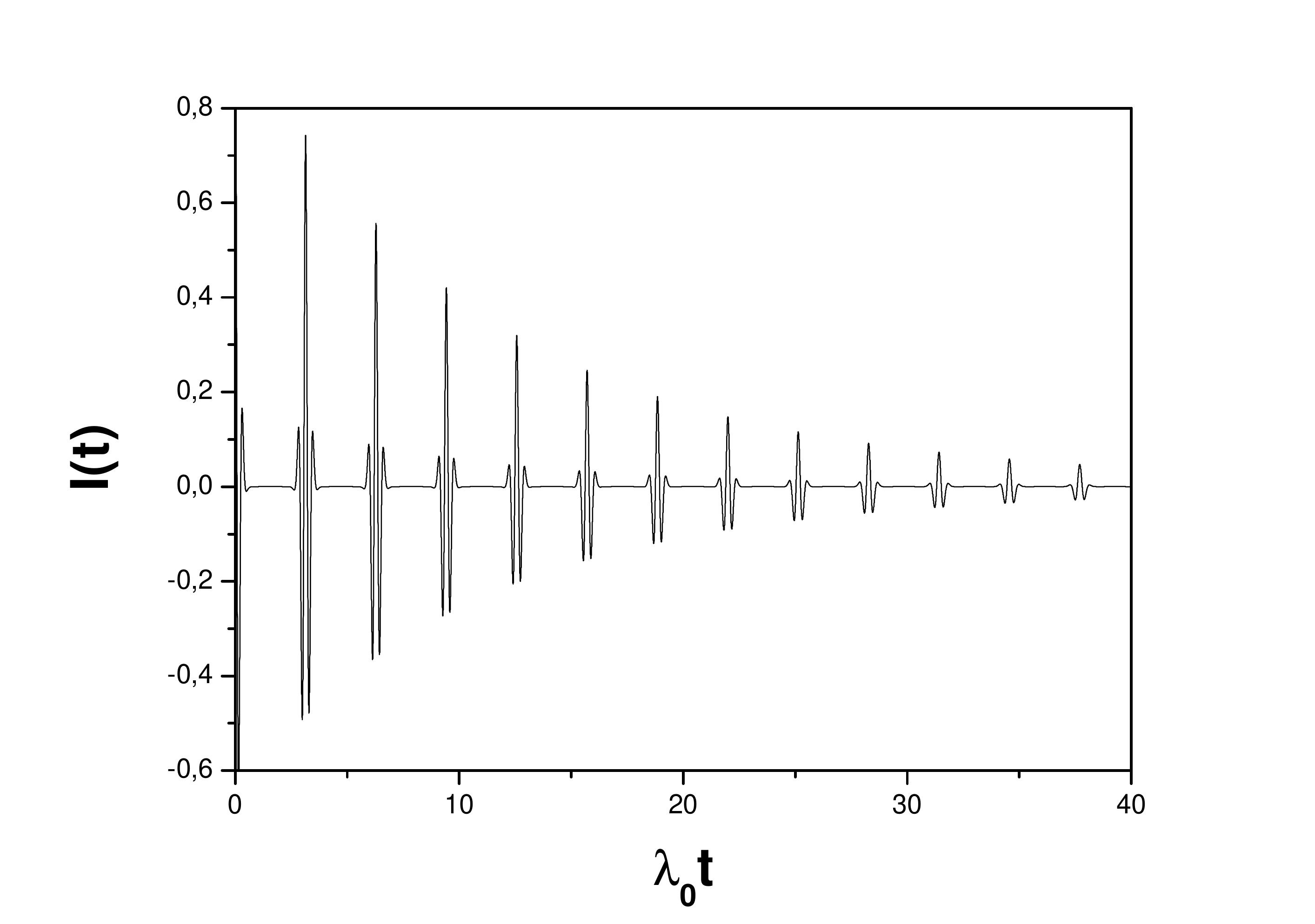} \label{i1c} }
\caption{Time evolution of the excitation inversion for different values of
the parameters $\protect\gamma (t)$ and $\protect\delta (t)$\ for: $%
\left\langle n\right\rangle =9$, $\protect\omega _{0}=\protect\omega %
_{c}=2000\protect\lambda _{0}$ (a) $\protect\gamma =0.001\protect\lambda %
_{0} $, $\protect\delta =0.0\protect\lambda _{0}$ and $f(t)=0;$ (b) $\protect%
\gamma =0.001\protect\lambda _{0}$, $\protect\delta =0.001\protect\lambda %
_{0}$ and $f(t)=0;$ (c) $\protect\gamma =0.001\protect\lambda _{0}$, $%
\protect\delta =0.005\protect\lambda _{0}$ and $f(t)=0.$}
\label{inv1}
\end{figure}
\begin{figure}[h!tb]
\centering
\subfigure(a){\includegraphics[height=4cm]{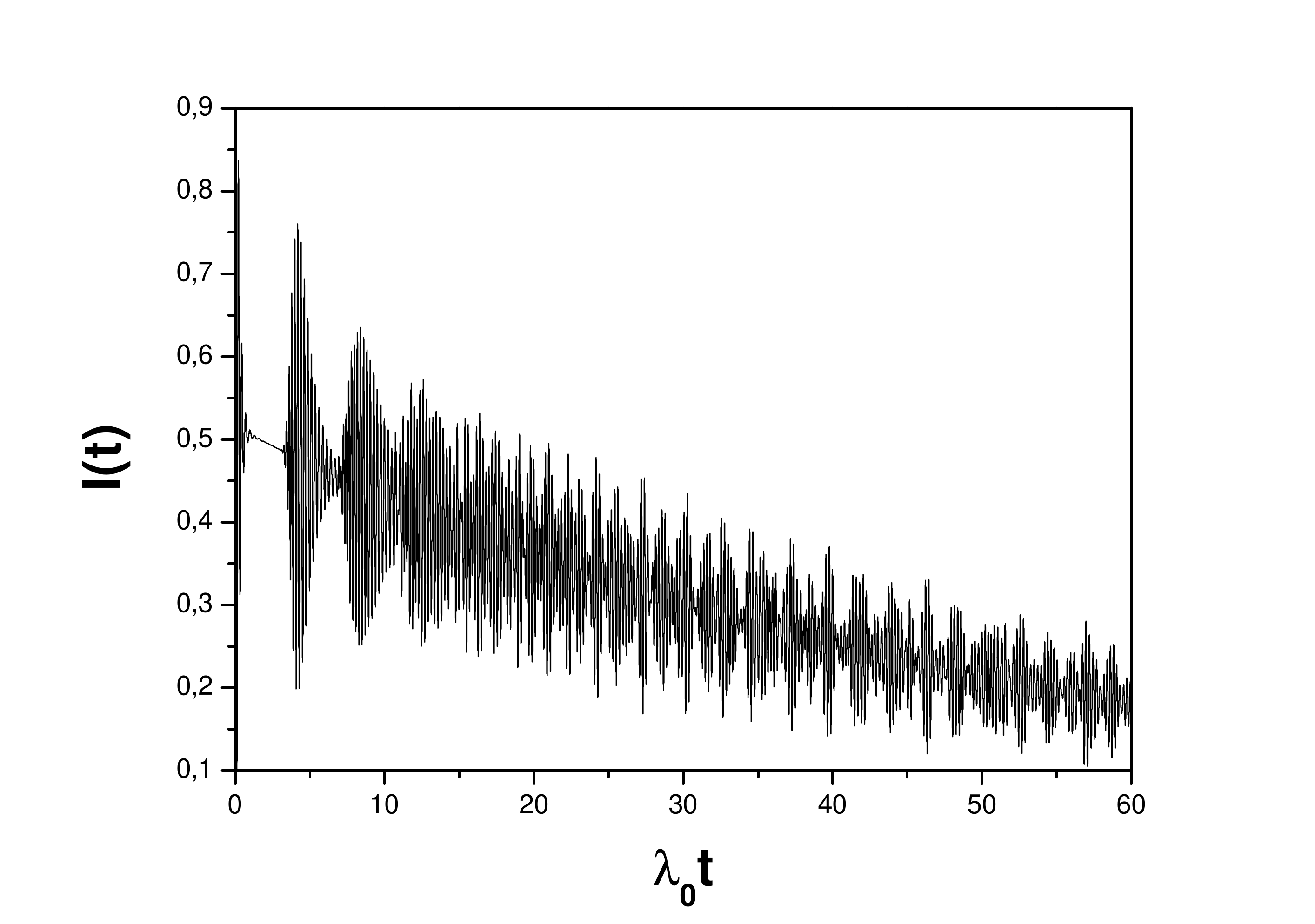} \label{i2a} } \quad 
\subfigure(b){\includegraphics[height=4cm]{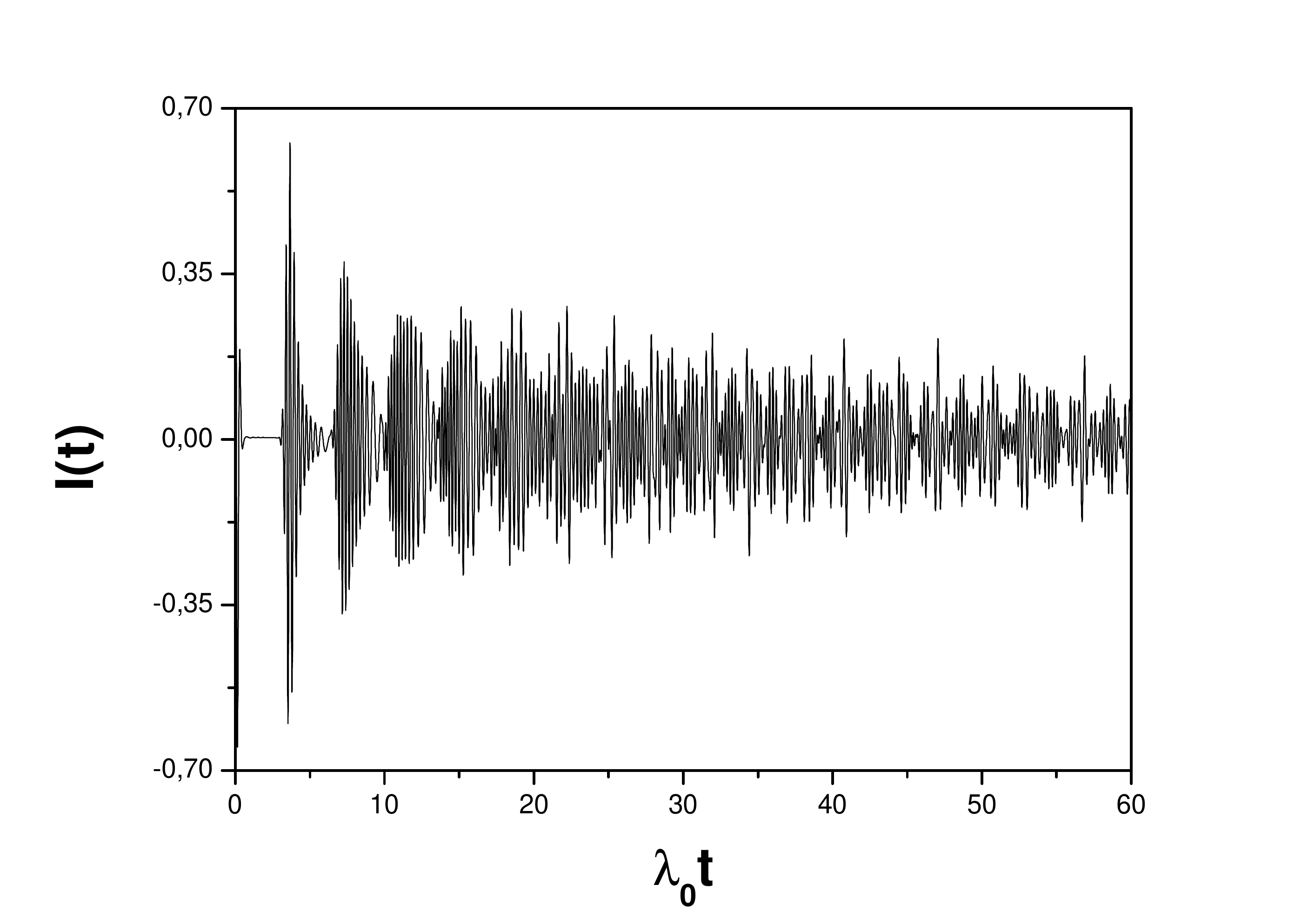} \label{i2b} } \quad 
\subfigure(c){\includegraphics[height=4cm]{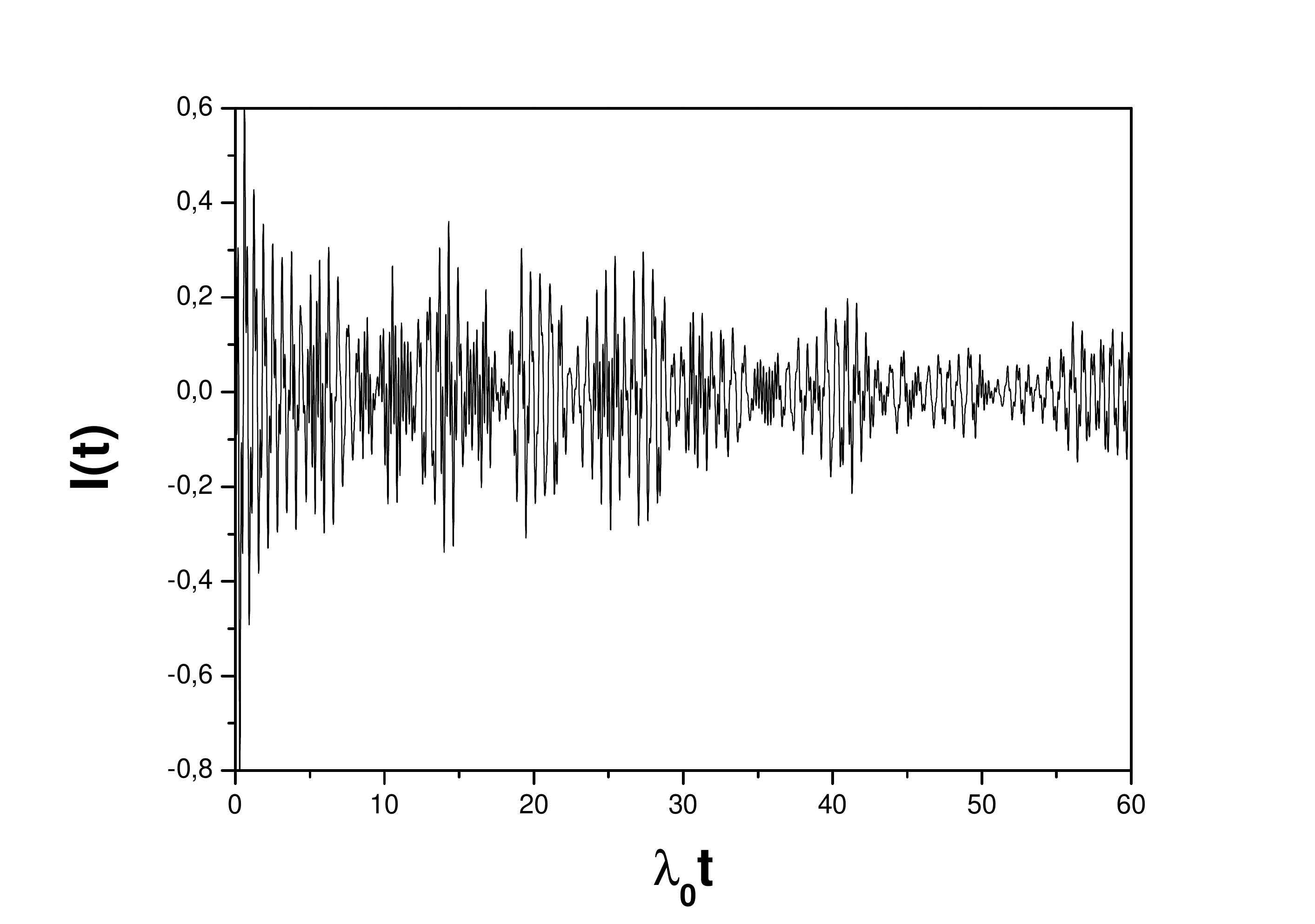} \label{i2c} }
\caption{Time evolution of the excitation inversion for different values of
the parameters $\protect\gamma (t)$ and $\protect\delta (t)$\ for: $%
\left\langle n\right\rangle =9$, $\protect\omega _{0}=\protect\omega %
_{c}=2000\protect\lambda _{0},$ $\protect\gamma =0.001\protect\lambda _{0}$, 
$\protect\delta =0.001\protect\lambda _{0}$ (a) $f(t)=20\protect\lambda %
_{0}; $ (b) $\protect\eta =20\protect\lambda _{0}$ and $\protect\omega %
\prime =\protect\lambda _{0};$ (c) $\protect\eta =20\protect\lambda _{0}$
and $\protect\omega \prime =20\protect\lambda _{0};$}
\label{inv2}
\end{figure}

Next we consider the time evolution of the \textit{excitation inversion }($%
EI $) for different values of the coefficients $\gamma (t)$ and $\delta (t)$%
. In the resonant case, $f(t)=0$, we have fixed\ again the \textbf{NR} with
average number of excitations $\left\langle \hat{n}\right\rangle =|\alpha
|^{2}=9$ and $\omega _{0}=\omega _{c}=2000\lambda _{0}$\ to calculate the $%
EI $ of the \textbf{CPB} (cf. Figs. \ref{inv1} (a), \ref{inv1} (b) and \ref%
{inv1} (c) ); we see in these three cases the $EI$ exhibiting similar
collapse-revival effects, but different amplitudes; these amplitudes
diminishes for larger values of the parameters $\gamma (t)$, $\delta (t)$.
In case of detuning, with $f(t)=\Delta =const.$, $\Delta \ll \omega
_{c},\omega _{0}$, Fig. \ref{inv2} (a) shows absence of the collapse-revival
effect; indeed even the $EI$ is also absent since these oscillations of \
the \textbf{NR} excitation characterize no inversion, as explained in the
``alert" below. On the other hand, for $f(t)=\eta \sin (\omega \prime t)$
the Fig. \ref{inv2} (b) stands for a variable detuning with maximum value $%
\eta $, equal to the value of the fixed value $\eta =\Delta =20$, Fig. \ref%
{inv2} (a). For small times the system presents the collapse-revival effect,
whereas for large times this effect vanishes, with concomitant attenuation
in the $EI$. In the \ref{inv2} (c) we use a large value for $\omega \prime $
and the same maximum value $\eta =20$ employed in the Fig. \ref{inv2} (b):
here we observe no occurrence of the collapse-revival effect, whereas the
amplitude of oscillations decreases for long times which is due to the
losses affecting both subsystems. This later result differs from the case $%
f(t)=\Delta $, as shown in Figs. \ref{inv1} (a), \ref{inv1} (b) and \ref%
{inv1} (c). Alert: two different properties must be distinguished in the
Fig. \ref{inv2} (a): one of them is the excitation inversion, the other is
the oscillation of excitation. The first is identified when the oscillations
changes the signs $(+)\longleftrightarrow (-)$ successively; the second
neglects these signs. Ignoring these aspects can confuse fluctuations with
inversion. For example, Fig. \ref{inv2} (a) shows oscillations, but not
inversion; then, what is really shown in this figure is a collapse of
oscillations, not a collapse of the excitation inversion.

\section{Conclusion}

We have studied the dynamical properties (entropy and excitation inversion)
of an interacting system composed by a \textbf{CPB} and a quantized \textbf{%
NR}. We have assumed the \textbf{CPB} initially in its excited state $%
|e\rangle $ and the \textbf{NR} initially prepared in a coherent state $%
|\alpha \rangle $. We also assumed the whole system \textbf{\textbf{CPB }- 
\textbf{NR}} described by the $BS$ model via single excitation transition.
Concerning the entropy, which is connected with entanglement (mixture) of
states, we have studied its time evolution in the presence of loss in both
subsystems, for the resonant case $(\Delta =0)$ and nonresonant case. The
influence upon the entropy of a (sinusoidal) time dependent frequency, via
the \textbf{NR} and coupling, was also considered. The EI was also
investigated under the same conditions. The inclusion of losses in the 
\textbf{CPB} and \textbf{NR} turns this scenario more realistic and the time
dependence of the coupling $\lambda (t)$\ and also the \textbf{NR}\
frequency $\omega (t)$\ make our results closer to experimental conditions.
The results drastically differ from those obtained in the resonant case. The
following scenarios were considered: $(i)$\textit{\ }the resonant case $(f=0)
$; $(ii)$ the nonresonant case, with fixed detuning, $f=\Delta \neq 0$, and $%
(iii)$ nonresonant case with time dependent detuning, $f(t)=\eta \sin
(\omega \prime t)$. An interesting result emerges: for a fixed detuning the
collapse-revival effect does not occurs, the same being true for the $EI$\
since the system oscillates around a value that differs from zero (cf. Fig. %
\ref{inv2} (a)) as also alerted before. However, it is surprising that in
the case $f(t)=\eta \sin (\omega \prime t)$, with the same conditions
assumed for fixed detuning, we can see the $EI$ effect remaining, even in
the presence of decay (cf. Fig. \ref{inv2} (b), Fig. \ref{inv2} (c)). This
behavior is not shown with fixed detuning. In summary, we have shown that
the use of a (time-dependent) modified $BS$ model, which is extended to a
more realistic scenario where the influence of the losses is considered, new
interesting findings emerge. They also indicate that it is possible to
perform a dynamic control of the system properties by changing the
parameters involved. Convenient choices of the frequency modulation can be
made to manipulate environmental noisy and inaccuracies, including potential
applications in the dynamical control of quantum information processes. We
hope that these results can offer a reference to put the issue with force.

\section{Acknowledgments}

We thank the CNPq and FAPEG for the partial supports and the UFG --
LCC-IF/UFG.

\end{document}